\documentclass[sigconf, screen]{acmart}
\usepackage{soul}
\usepackage{hyperref}
\usepackage{multirow}
\usepackage{fancybox}
\usepackage{enumitem}
\usepackage{graphicx}
\usepackage{balance}
\usepackage{color}
\usepackage{float}
\usepackage{xcolor}
\usepackage{array}
\usepackage{amsmath}
\usepackage{xspace}
\usepackage{url}
\usepackage{tikz}
\usepackage{caption}
\usepackage{pgfplots}
\usepackage{listings}
\usepackage{color}
\usepackage{graphicx}
\definecolor{dkgreen}{rgb}{0,0.6,0}
\definecolor{gray}{rgb}{0.5,0.5,0.5}
\definecolor{mauve}{rgb}{0.58,0,0.82}
\pgfplotsset{compat=1.16}
\usepackage{pgf-pie}
\usetikzlibrary{pgfplots.statistics,calc}
\usepackage{algorithm}
\usepackage{algpseudocode}
\usepackage{subcaption}
\usepackage{listings}

\usepackage{tcolorbox}

\makeatletter
\newcommand{\mybox}[1]{%
	\setbox0=\hbox{#1}%
	\setlength{\@tempdima}{\dimexpr\wd0+13pt}%
	\begin{tcolorbox}[boxrule=0.5pt, colback=white, arc=4pt,
		left=6pt,right=6pt,top=6pt,bottom=6pt,boxsep=0pt]
		#1
	\end{tcolorbox}
}

\definecolor{codegreen}{rgb}{0,0.6,0}
\definecolor{codegray}{rgb}{0.5,0.5,0.5}
\definecolor{codepurple}{rgb}{0.58,0,0.82}
\definecolor{backcolour}{rgb}{0.95,0.95,0.92}

\lstdefinestyle{mystyle}{
  language=Java,
  aboveskip=3mm,
  showstringspaces=false,
  columns=flexible,
  numbers=none,
  backgroundcolor=\color{backcolour},
  commentstyle=\color{codegreen},
 keywordstyle=\color{magenta},
    numberstyle=\tiny\color{codegray},
    stringstyle=\color{codepurple},
    basicstyle=\small\ttfamily,
    breakatwhitespace=false,         
    breaklines=false,                 
    captionpos=b,                    
    keepspaces=false,                 
    numbersep=5pt,                  
    showspaces=false,                
    showstringspaces=false,
    showtabs=false,                  
    tabsize=2,
    escapeinside=``
}
\definecolor{nima2}{RGB}{1.0, 0.49, 0.0}
\definecolor{songcolor}{RGB}{191,191,191}
\definecolor{nimacolor}{RGB}{0.13, 0.67, 0.8}
\definecolor{aruncolor}{RGB}{51,255,51}

\AtBeginDocument{%
  \providecommand\BibTeX{{%
    \normalfont B\kern-0.5em{\scshape i\kern-0.25em b}\kern-0.8em\TeX}}}

\setcopyright{acmlicensed}
\copyrightyear{2024}
\acmYear{2024}
\acmDOI{XXXXXXX.XXXXXXX}

\acmConference[ISSTA 2024]{ACM SIGSOFT International Symposium on Software Testing and Analysis}{16-20 September, 2024}{Vienna, Austria}
\acmBooktitle{ISSTA '24: ACM SIGSOFT International Symposium on Software Testing and Analysis,
 September 16--20, 2024, Vienna, Austria} 
\acmISBN{978-1-4503-XXXX-X/18/06}

\begin{document}

\title{Domain Adaptation for Code Model-based Unit Test Case Generation}

\author{Jiho Shin}
\email{jihoshin@yorku.ca}
\affiliation{%
  \institution{York University}
  \streetaddress{4700 Keele St.}
  \city{Toronto}
  \state{Ontario}
  \country{Canada}
  \postcode{M3J 1P3}
}
\author{Sepehr Hashtroudi}
\email{sepehr.pourabolfathh@ucalgary.ca}
\affiliation{
  \institution{University of Calgary}
  \streetaddress{2500 University Dr NW}
  \city{Calgary}
  \state{Alberta}
  \country{Canada}
  \postcode{T2N 1N4}
}

\author{Hadi Hemmati}
\email{hemmati@yorku.ca}
\affiliation{%
  \institution{York University}
  \streetaddress{4700 Keele St.}
  \city{Toronto}
  \state{Ontario}
  \country{Canada}
  \postcode{M3J 1P3}
}

\author{Song Wang}
\email{wangsong@yorku.ca}
\affiliation{%
  \institution{York University}
  \streetaddress{4700 Keele St.}
  \city{Toronto}
  \state{Ontario}
  \country{Canada}
  \postcode{M3J 1P3}
}
\renewcommand{\shortauthors}{Shin et al.}

\begin{abstract}
Recently, deep learning-based test case generation approaches have been proposed to automate the generation of unit test cases.
In this study, we leverage Transformer-based code models to generate unit tests with the help of Domain Adaptation (DA) at a project level. Specifically, we use \emph{CodeT5}, a relatively small language model trained on source code data, and fine-tune it on the test generation task.
Then, we apply domain adaptation to each target project data to learn project-specific knowledge (project-level DA).
We use the \emph{Methods2test} dataset to fine-tune \emph{CodeT5} for the test generation task and the \emph{Defects4j} dataset for project-level domain adaptation and evaluation. 
We compare our approach with (a) \emph{CodeT5} fine-tuned on the test generation without DA, (b) the \emph{A3Test} tool, and (c) \emph{GPT-4} on five projects from the \emph{Defects4j} dataset.
The results show that tests generated using DA can increase the line coverage by 18.62\%, 19.88\%, and 18.02\% {and mutation score by 16.45\%, 16.01\%, and 12.99\%} compared to the above (a), (b), and (c) baselines, respectively.
The overall results show consistent improvements in metrics such as {parse rate, compile rate,} BLEU, and CodeBLEU.
In addition, we show that our approach can be seen as a complementary solution alongside existing search-based test generation tools such as \emph{EvoSuite}, to increase the overall coverage and mutation scores with an average of 34.42\% and 6.8\%, for line coverage and mutation score, respectively.
\end{abstract}



\received{15 December 2023}
\received[revised]{9 February 2024}
\received[accepted]{2 March 2024}

\maketitle

\section{Introduction}
\label{sec:intro}
Code models that are pre-trained on a large corpus of source code have been introduced to automate numerous software development tasks such as comment generation, code translation, and code generation \cite{shin2021survey,shin2023prompt,kanade2020learning,ahmad2021unified}. 
Among these downstream tasks, unit test generation, which can be seen as a neural machine translation task, has started gaining its spotlight recently \cite{tufano2020unit, alagarsamy2023a3test}. 

There are several reasons for the challenges in unit test case generation: (a) The robustness of the code generation model is more challenging to achieve, as slight miss generation would lead to an error.
Unit test case generation, in particular, might be more challenging than regular code generation as test cases tend to have more minor differences between code. For example, a line of assertion statements or a couple of statements to instantiate objects might drive the program into an interesting and testable state.
(b) Properly evaluating the generated test cases requires executing the generated tests to calculate test adequacy metrics, which is time-consuming and typically requires non-trivial manual labor, e.g., resolving dependencies. 
(c) Domain shift problem~\cite{zirak2022improving} occurs when the pre-trained models cannot transfer their code knowledge to a new target project due to different code distributions in various domains of projects.

Despite these shortcomings, test case generation based on deep neural code models has advantages.
The generated tests from neural models are similar since the models are trained on human-written code. Therefore, they are more readable and maintainable than the alternative automatically generated test cases.
As previous literature suggests~\cite {tufano2020unit}, developers prefer neural model-generated tests over other automatically created test cases since they are more readable and understandable.
They also target different faults (the same as those targeted by the developer-written tests) compared to tests generated by, e.g., search-based approaches, which usually focus on maximizing code coverage. 

To address the shortcomings of pre-trained code models for test case generation, i.e., low performance, insufficient evaluation, and domain shift, we propose a simple yet novel technique by adopting two different levels of fine-tuning/domain adaptation: task and project. In our approach, first, we fine-tune the \emph{CodeT5} pre-trained model with a task-specific dataset to customize the model for generating unit test cases, given a method under test. Then, we apply domain adaptation with the project-specific dataset to learn the proper code knowledge and create higher-quality test cases for mitigating the impact of the domain shift problem. We also conduct a more thorough investigation by evaluating test adequacy and textual similarity metrics to address the insufficient evaluation problem. 
Regardless of the simplicity of the idea, we note that this approach is 1) novel and 2) effective as it enables the relatively smaller model (\emph{CodeT5} with 220M parameters) to outperform much bigger models (\emph{GPT-4} with 1.76T).
Our framework uses automated post-processing of simple heuristics to mitigate compilability/executability issues.
We use the \emph{Methods2test} dataset \cite{tufano2022methods2test} for fine-tuning the test case generation task.
We apply domain adaptation to the models by leveraging human-written unit test cases for each project.
For evaluation and domain adaptation, we use the \emph{Defects4j} dataset \cite{just2014defects4j}.
We compare the effectiveness of our approach with and without domain adaptation.
We also investigate two other baselines, namely \emph{GPT-4} (the largest and the state-of-the-art LLM) and \emph{A3Test} (state-of-the-art neural test case generation method which exploits task-knowledge domain adaptation).
Our model with project-level domain adaptation outperforms all the baselines on {all the studied metrics, except for the parse rate of \emph{GPT-4}.}
Furthermore, our approach can be used alongside search-based test generation to increase their line coverage and mutation score. 

We show that using domain adaptation, we can improve the line coverage with an average of 18.62\%, 19.88\%, and 18.02\% {and mutation score by 16.45\%, 16.01\%, and 12.99\%} compared to \emph{CodeT5} without DA, \emph{A3Test}, and \emph{GPT-4} baselines, respectively.
We also show that our approach can increase the overall coverage and mutation scores of \emph{EvoSuite} when used alongside each other, with an average of 34.42\% and 6.8\% for line coverage and mutation score, respectively. 

In summary, our main contributions are as follows:
\begin{enumerate}
    \item We propose a line-level neural test case generation framework leveraging domain adaptation, which creates high-quality unit test cases {(compilable, similar to human-written, and test-adequate)}. 
    \item We conducted an empirical study on \emph{Defects4j} benchmark dataset~\cite{just2014defects4j}, which shows our approach improves the performance of the most related work \emph{AthenaTest}, \emph{A3Test}, and \emph{GPT-4}) from the literature.
    \item  We also show that our approach can cover lines that neither developer-written tests nor a baseline search-based testing tool can cover. We also showed that we can kill new mutants compared to the search-based tools.  
    \item Unlike most related work, we execute the generated test cases and evaluate them with proper test adequacy metrics (i.e., code coverage and mutation score), which require much more effort to calculate compared to BLEU/CodeBLEU. We also report the BLEU and CodeBLEU scores, which are much used in the literature for automated evaluation metrics.
\end{enumerate}

The code for our proposed approach and the experiment's scripts and raw data are publicly available for replication\footnote{\url{https://github.com/shinjh0849/unit_tc_generation}}. 

We organized the rest of this paper as follows.
Section~\ref{sec:bg} introduces the background of neural models for code and unit test generation. 
Section~\ref{sec:app} presents the approach of our test case generation framework.
Section~\ref{sec:settings} shows the experimental setup. 
Section~\ref{sec:evaluation} presents the evaluation results. 
Section~\ref{sec:threats} discusses the possible threats in our study.
Section~\ref{sec:con} concludes this paper. 
\section{Background and Related Work}
\label{sec:bg}

\subsection{Search-based Software Testing}
In search-based software testing (SBST), the problem of test case generation is translated into an optimization problem over a test adequacy criterion such as code coverage \cite{mcminn2004search}.
For instance, \emph{EvoSuite} \cite{fraser2011evosuite} is an SBST tool that generates test cases to optimize statement or branch coverage of the generated tests.
It uses a genetic algorithm to evolve a test suite toward a higher quality set (more coverage with minimum tests).
While SBST tools have shown great effectiveness, studies report their limitation in understandability or readability \cite{roy2020deeptc, grano2018empirical, daka2015modeling}, quality \cite{grano2019scented, palomba2016automatic}, and their performance in detecting actual bugs from the generated unit test cases. \cite{pinto2010multi, bacchelli2008effectiveness}

\subsection{Domain Adaption}
Domain adaptation is a technique for modifying a model trained on one domain to perform well on a different but related domain.
The goal is to leverage the knowledge gained from the source domain to improve the performance of the target domain, mainly when the target domain has limited labeled data. Domain adaptation is a type of transfer learning which aims to transfer knowledge from one task to another.

Nam et al. \cite{nam2013transfer} proposed a novel transfer defect learning approach, \emph{TCA+}, which applies a transfer learning technique to reduce the data distribution difference between source and target projects for cross-project defect prediction. \emph{TCA+} also selects a suitable normalization option based on the similarity of data set characteristics between the source and target projects and significantly improves prediction performance.
Patel et al. \cite{patel2015visual} did a survey about domain adaptation methods for visual recognition.
The paper discusses the challenges, assumptions, and formulations of domain adaptation and categorizes the existing methods into feature augmentation, feature transformation, parameter adaptation, dictionary learning, and others. It also highlights the advantages and limitations of each category and identifies some promising directions for future research in this field.
Farahani et al. \cite{farahani2021brief} have briefly reviewed domain adaptation. It introduces the main categories, challenges, and domain adaptation approaches, focusing on unsupervised domain adaptation.
Zirak et al. \cite{zirak2022improving} propose a domain adaptation framework for automated program repair (APR) models that can improve their effectiveness on new and different projects. The framework uses three methods: full fine-tuning, tuning with lightweight adapter layers, and curriculum learning. It also employs a data synthesis method to create artificial bugs for zero-shot learning.

\subsection{Neural Models for Unit Test Generation}
Deep neural models of code for unit test case generation are limited and relatively new.
They can be grouped into two categories, i.e.,
test oracle generation and unit test case generation. 

\subsubsection{Test Oracle Generation}
Test oracle generation aims to generate oracles, e.g., meaningful assertion statements when the focal context (method under test together with its class information, i.e., class method signature and class fields) and the corresponding test prefix are given \cite{shin2023assessing}.
Test prefixes are statements in a unit test case with the oracles (assertion statements, try-catch clause, etc.) removed.
Test prefixes drive the program into a desired testable state.
In general, the problem of oracle generation is a subset of the whole test case generation.

\emph{ATLAS} (\underline{A}u\underline{T}omatic \underline{L}earning of \underline{A}ssert \underline{S}tatements) \cite{watson2020learning} is the first to utilize deep neural models for assertion generation. 
They could generate assertions with the BLEU-4 score of 61.85\%.
Yu et al. \cite{yu2022automated} introduced an approach to integrate information retrieval techniques, using Jaccard coefficient \cite{tanimoto1958elementary}, Overlap \cite{wiki2023overlap}, and Dice coefficient \cite{dice1945measures} with the deep neural approach \emph{ATLAS}.
With their approach, they could boost the BLEU score up to 78.86\%.
\emph{TOGA} (a neural method for \underline{T}est \underline{O}racle \underline{G}ener\underline{A}tion) \cite{dinella2022toga} was proposed to use a unified transformer-based neural model to generate both try-catch clause and assertion statements for unit test case oracles.
For generating the try-catch clause, they had 86\% of exact match accuracy and 69\% for assertion statements.
Tufano et al.~\cite{tufano2022generating} proposed to apply the \emph{BART} pre-training model trained with natural language and source code corpus and then fine-tune on \emph{ATLAS} dataset. 
They achieved an exact match accuracy of 62.47\% with a beam size of one.

The main difference between our work and test oracle generation is that test oracle generation models only focus on the oracle part of the test case.
Generating test prefixes is a non-trivial task, which calls into the need to generate whole unit test cases.

\subsubsection{Unit Test Case Generation}
There have not been many studies related to automating the generation of whole test cases.
Liu et al. \cite{liu2017automatic} exploited deep learning models to generate relevant text inputs to test user interfaces for mobile applications.
Saes \cite{saes2018unit} generated a test suite for Java projects by identifying the connections between focal methods and their corresponding tests.
They have gathered more than 780K pairs of focal and test methods utilizing the JUnit testing framework from GitHub.
They could generate test cases with a parsability of 86.69\%.
However, they did not evaluate how correct or effective the generated test cases were in identifying bugs or covering code.
Tufano et al. \cite{tufano2020unit} proposed \emph{AthenaTest} which exploits the \emph{BART} pre-trained model on both natural language and source code corpora then fine-tune on \emph{Methods2Test} \cite{tufano2022methods2test} dataset, to generate whole unit test cases when a focal method and its context is given.
They have found that their method could correctly test 43\% focal methods, with 16\% of the candidates being correct.
Alagarsamy et al. \cite{alagarsamy2023a3test} proposed \emph{A3Test}, which is a test case generation approach that is augmented by a test oracle generation task and includes a mechanism to verify naming consistency and test signatures. It performs domain adaptation at a task level, i.e., test oracle generation task to whole test case generation task, achieving more correct test cases and method coverage than \emph{AthenaTest}. 
{
Lemieux et al. \cite{lemieux2023codamosa} proposed \emph{CodaMOSA}, an SBST approach that leverages LLM for escaping the coverage plateau for Python code bases.
Schafer et al. \cite{schafer2023empirical} proposed \emph{TestPilot}, a test cases generation approach that leverages LLM, usage examples mined from package documentation, and error logs for npm packages (JavaScript).
Yuan et al. \cite{yuan2023no} proposed \emph{ChatTester}, which is a LLM-based test case generation model that exploits \emph{ChatGPT} and iterative generate-and-validate prompt engineering strategy with execution feedback.
Nie et al. \cite{nie2023learning} proposed \emph{TeCo}, a deep encoder-decoder test completion model that learns different levels of code semantics and re-ranking by execution. A test completion model generates the following statement of a unit test case when the previous line and method under test are given.
}
Our study continues in this direction and proposes domain adaptation at a project level to improve \emph{AthenaTest} and \emph{A3Test} as our most related work.
Unlike these papers, we evaluate based on classic software testing criteria (i.e., code coverage and mutation testing). Most existing approaches only report BLUE scores or similar NLP-based metrics that do not correlate with the effectiveness (adequacy) of the generated test cases.
Although there is literature regarding test case generation, it has shown that we still have challenges in generating correct and effective test cases that reveal bugs for practical usage.

\section{Test Case Generation with Project Level Domain Adaptation}
\label{sec:app}
\begin{figure}[t!]
  \centering
  \includegraphics[width=0.9\linewidth]{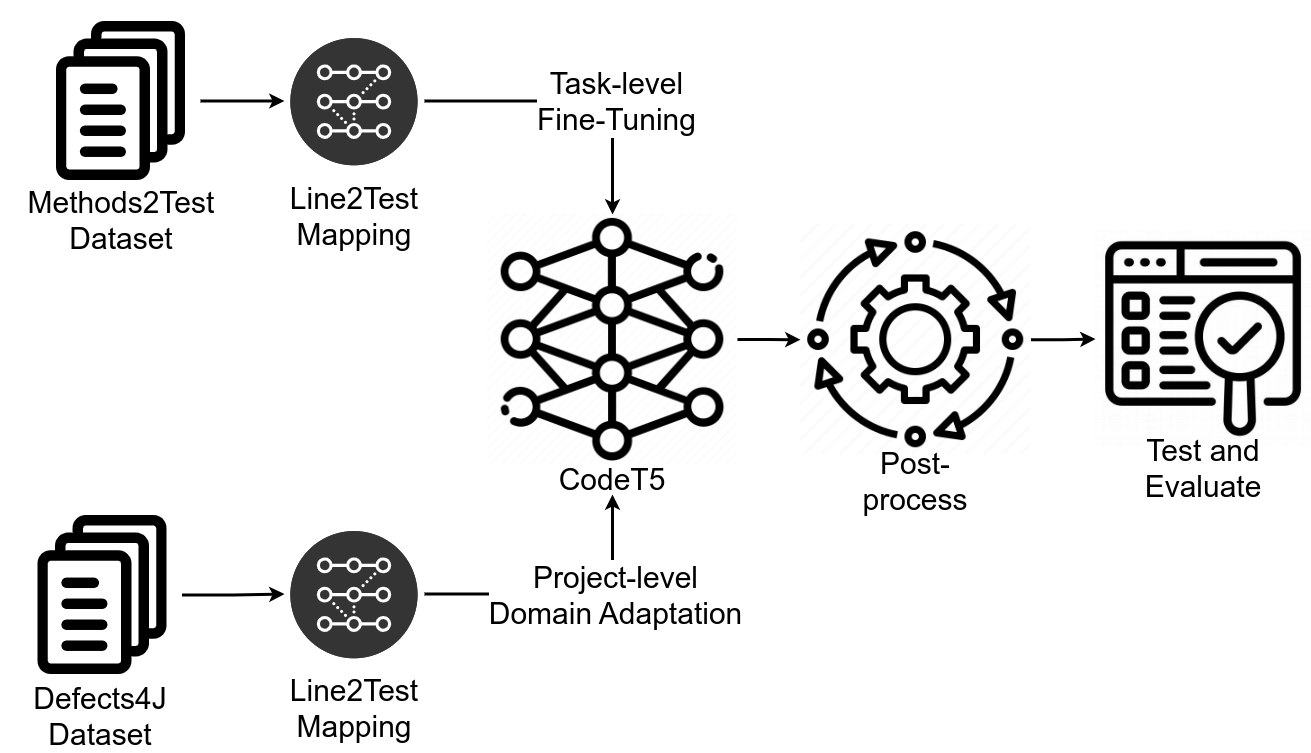}
  \caption{Overview of our approach.}
  \label{fig:overview}
\end{figure}

Figure \ref{fig:overview} illustrates our proposed test case generation framework.
Our approach contains two major steps: (a) fine-tuning the \emph{CodeT5} model on a task-level dataset and then (b) applying domain adaptation (DA) on a project-level dataset. 
The following sections explain the two steps in detail.
The framework aims to generate high-quality test cases with adequate test efficacy learned from developer-written test cases. 

\subsection{Fine-tuning on Test Case Generation Task}
\label{sec:noda_approach}
Our framework assumes the project under test has an initial test suite.
We aim to improve the project by generating new tests using code models.
Although we use developer-written test suites as our initial set, they can also be automatically generated (e.g., using ChatGPT).
We explain each option's limitation in the threats to the validity section. 

The first step is to create a \textbf{coverage database} from the existing test suite.
We use the line-level coverage in our framework and evaluation for simplicity.
However, this can be extended to other code metrics or mutation scores.
The coverage database keeps the information on which unit test covers which lines of source code.
In the next step, our \textbf{line2test mapping} approach converts the coverage data to map each line in the source code to its covering tests.
\textbf{Line2test mapping} extracts the classpath of all test cases.

Next, we fine-tune the \emph{CodeT5} model on the ``test case generation'' downstream task.
We fine-tune the \emph{CodeT5} model since they are not specifically trained on test generation tasks.
We use ``conditional code generation'' proposed by the original \emph{CodeT5} paper to optimize the model for fine-tuning.
Conditional code generation generates code similar to conventional natural language processing.
They adapt the conventional Sequence-to-Sequence framework for learning the task-specific data \cite{sutskever2014sequence}.

The dataset to fine-tune the code model for the test generation downstream task is the \emph{Methods2Test} dataset \cite{tufano2022methods2test} (henceforth \textbf{test generation data}).
It consists of tuples of a focal method, focal context (e.g. class name, class fields, public method signatures), and the associated unit test method.
Our framework maps between the input source method, the context of the input source method, and a test method that covers the source method. 

The final step in the framework is \textbf{post-processing} the test cases. 
Since the model accepts data input and output in one line, we replaced the new line ("\textbackslash n") characters with the ``[EOL]'' token. 
The first step in post-processing is to replace ``[EOL]'' with ``\textbackslash n''. 
Then, we make a list of generated test cases.
We add a unique number at the end of the duplicate test names to prevent compile errors due to duplicate definitions of the same test case. 
The next step is to select compilable tests because not all the model-generated tests are compilable. We include another step to automate the inclusion/exclusion process. 
Since compiling all the tests is time-consuming, we first use a Java parser to select the tests without syntax errors.
We use tree-sitter \cite{treesitter} for the implementation, a parser that supports multiple programming languages.
Since the parser does not compile the code, we can identify all parsing issues in a class in less than a second.  
Some generated tests are truncated due to the model's token limitation, thus will result in a parse error.
To make them parsable, we apply a simple bug-fix pattern to them.
These tests generally follow the same pattern, i.e., they do not match brackets or have a missing ``;'' at the end of the last line.
Using an automated script, we fix these issues by deleting the last line (usually, the test ends in the middle of a line) and adding a closing bracket.
Then, we parse the test again. If it is still not parsable, we add another closing bracket and re-parse.
After selecting the final parsable tests, we add each test to its corresponding test class to compile them in the project environment.
We use the classpath for each test from the \textbf{line2test mapping} step.
The test class has the required dependencies and other test helpers that the test case may need to be compilable.
We add each test case to the corresponding file and compile it individually to ensure it's also compilable.
Some tests may not pass this step for several reasons, such as calling undefined or unreachable objects or functions.
To fix these compile issues, we add the test cases to their corresponding test class file consisting of test helpers and other dependencies (saved in the line2test step). 

Finally, we remove all the developer-written tests and add each model-generated unit test to its corresponding test class.
Even with one test with a compilation error, the build would not succeed.
Thus, we add one test case at a time and compile the project using the \emph{Defects4j} framework.
If the test is compilable, we add it to the list of parsable and compilable tests; if it does not compile, we exclude it from our test set. 
Using this fully automated post-processing step, we now have a set of compilable model-generated test cases.

\subsection{Domain Adaptation on Project Specific Knowledge}
\label{sec:da_approach}
The main downside of a test case generation model is its inability to adapt to potential domain shifts when the model is inferring a new project. This phenomenon (domain shift) is not limited to the test case generation task and applies to all machine learning tasks. The structural difference in each project may cause a drastic change in the context generated by the framework making it harder for the model to generate compilable tests. We leverage the existing developer-written tests for each project to mitigate this threat. Usually, a well-maintained project already has a test suite covering most of the code. We use the existing test suite to generate a project-specific dataset for domain adaptation. Alternatively, one can start with a set of automatically generated test cases and further improve them with our approach.

As demonstrated in Figure \ref{fig:overview}, the first step for applying domain adaptation is to generate a dataset using developer-written tests.
Since a target line can be as simple as a return statement or an arithmetic operation of two variables, it is hard for a model to generate a meaningful test case.
To help the model generate meaningful unit tests, we append the target line as extra context/information.
Context extraction provides three different outputs.
The first output is a set of files identical to the files seen in the project source code structure, but in each file, instead of the full implementation, we only include the method names in that file. 
The second output is the same set of files, but it includes the method bodies.
We also save each method's initial and last line numbers in each file. 
The third output is all the other parts of the context for each class, consisting of the class name, signatures of the constructor methods, public variables and fields, all other method names, etc. 
This context design strictly follows the baseline work from the \emph{Methods2Test} dataset, which we used for fine-tuning downstream tasks.
After extracting the three context outputs, we iterate over the \textbf{Line2test mapping} and concatenate each line with its corresponding focal method and context. 

\begin{figure}[ht]
  \centering
  \includegraphics[width=\linewidth]{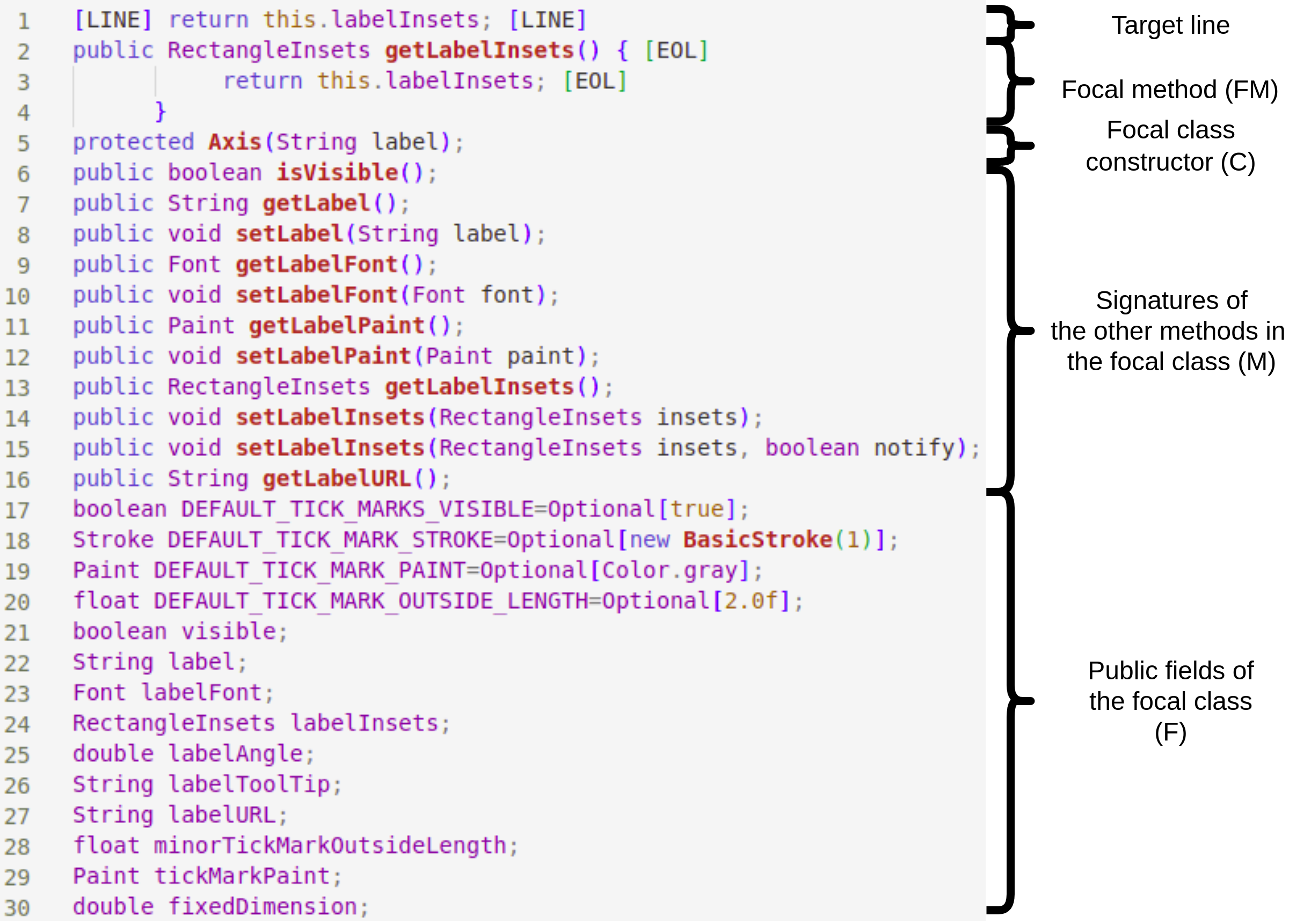}
  \caption{An example method and its context.}
  \label{fig:method_sample}
\end{figure}

An example of the input of the dataset is demonstrated in Figure \ref{fig:method_sample}.  
When lines are mapped to tests, we can end up with multiple test cases covering the same lines. 
We don't include all the covering test cases because the input data is the same, and the model will have difficulty optimizing if we provide different outputs for identical inputs. So, we select one test per line. 
We use the naming convention as a typical solution in the literature to map unit tests to source code. We have the test name and path in our \textbf{line2test mapping} and the class name to which the input line belongs. So, we search for the class name of the input line in different tests that have covered the line; if we have a match, we will select that test as the covering test for the input line. If no tests have the same class name as the input line, we include the first test in the list as the unit test. We select the first match as the mapped unit test if there are multiple matches.

After creating the dataset consisting of the input line, the context on the input side, and the corresponding test that covers the line, we use them to apply \emph{domain adaptation} to the fine-tuned model. The \emph{domain adaptation} enables the model to adapt to the new domain (project) and generate more accurate tests with a higher compilation ratio.
The domain adaptation technique we have used is training the model for a few epochs on the generated dataset (much less than a full training). Other options could be freezing the model and adding extra layers (which we did not see a significant benefit in performance from our preliminary experiment). However, this part of our approach can be easily replaced with other techniques in future work.

Note that the computation cost of domain adaptation per project is affordable since it is a one-time cost. If the project tends to change over time, we can re-train the model per sprint. Generating inferences requires significantly less time than classic test case generation approaches, i.e., SBST tools.
However, it is worth mentioning that SBST approaches do not require this non-trivial training time. So, deciding which approach is more time-effective is worth investigating for future work.
\section{Experiment Settings}
\label{sec:settings}
\subsection{Research Questions}
This study aims to evaluate the effectiveness of code models in generating unit test cases. In addition, we also study the effect of project-level domain adaptation. To address these objectives, we design and study the following three Research Questions (RQs):

\subsubsection{\textbf{RQ1 (Performance)}}
\label{RQ:1}
How effectively does \emph{CodeT5} generate unit test cases without domain adaptation?

\textbf{Motivation:}
Code models such as \emph{CodeBERT} and \emph{CodeT5} have been successfully applied to automated software engineering tasks such as comment generation, defect prediction, and program repair.
However, not many studies apply them to test case generation tasks. The few existing studies, such as \cite{tufano2022generating,alagarsamy2023a3test}, mainly rely on the static evaluation of the generated test cases using metrics such as BLEU. The problem with this approach is that the generated test may be (a) uncompilable and (b) not effective since they are measured by execution. Therefore, we replicate a state-of-the-art study in this domain and adequately evaluate them by running test cases and calculating their test adequacy metrics. We take \emph{AthenaTest} \cite{tufano2020unit} as our baseline.
However, they have not published their model publicly, and our attempt to access it privately was also unsuccessful (due to confidentiality). 
Also, they have not reported the test adequacy metrics (e.g., line coverage and mutation score). So we replicated their work the best we could, using a similar model \emph{CodeT5} \cite{wang2021codet5} and evaluated it on our dataset.

\subsubsection{\textbf{RQ2 (Effectiveness)}} 
How effective is project-level domain adaptation in improving the generation quality of test cases using \emph{CodeT5}?

\textbf{Motivation:}
\label{rq2:motiv} As we show in Section \ref{sec:rq1_result}, the results of RQ1 are not promising. One potential explanation is the inability of the models to learn project-specific patterns or knowledge. It is a typical problem in most software engineering tasks where ``domain shift'' is extreme when a trained model is applied on a brand new project \cite{zirak2022improving}. 
In most well-maintained real-world projects, the code base already includes developer-written test cases that may cover most of the code.
In RQ2, we propose to leverage this data to adapt the domain in RQ1 results to each project.
In this RQ, we compare the results to other baselines, namely, \emph{GPT-4} and \emph{A3Test}.
We added \emph{GPT-4} as a method to represent the current state-of-the-art LLM and \emph{A3Test} as the state-of-the-art technique specially designed for unit test case generation exploiting domain adaptation from test oracle generation task. 
 
\subsubsection{\textbf{RQ3 (Impact)}} Can our proposed test generation approach augment search-based test case generation?

\textbf{Motivation:}\label{RQ3:Motivation}
Given that search-based test case generation approaches are state-of-the-art techniques for test case generation, in RQ3, we compare our approach with \emph{EvoSuite} as a well-known and influential search-base approach\cite{fraser2016evosuite}. 
One of the motivations for using our approach compared to a technique like \emph{EvoSuite} is the speed of test generation. Given that test generation in our approach only requires one single ``inference'', the run-time per test case should be substantially lower than search-based baselines. On the other hand, those techniques do not need a lengthy (one-time) training phase beforehand. 
Our objective for the proposed test generation framework is not to replace search-based approaches but to complement them. We argue that search-based approaches and our Transformer-based methods can target different types of tests to generate. Therefore, using both tools together will be most beneficial. In particular, we envision applying \emph{EvoSuite} once per class to reach a certain level of coverage. Then, running our models (domain adapted on the developer-written tests), we can generate extra test cases covering some new lines on top of the coverage achieved by \emph{EvoSuite} and developer-written tests combined. After the initial round of test generation, our model can be run after each small commit to cover those few new source code lines. We will discuss the complementary nature of the two approaches in Section \ref{eval:rq3}. We also compare the mutation score of both methods and report how many new mutants our method can kill compared to \emph{EvoSuite}.

\subsection{Datasets}
\label{dataset}
We use two datasets in this study namely \emph{Methods2Test} and \emph{Defects4j}.
The \emph{Methods2Test} dataset is used to fine-tune our model for test generation downstream tasks.
\emph{Defects4j} projects are then used to adapt the fine-tuned model at a project level and test their final output per project. In the following, we provide more details about each of these datasets.

\textbf{Methods2Test~\cite{tufano2020unit}} is the ``test generation data'' in Figure \ref{fig:overview}, which consists of Java methods mapped to their corresponding focal methods.
It was built on data from 9,410 unique repositories (91,385 original repositories analyzed). 
The dataset consists of 780,944 instances, divided into training (80\%), validation (10\%), and test (10\%) sets.
Since running all the tests, getting the coverage, and mapping the tests to the methods they cover require a lot of manual work and execution time, they have used the naming convention as a heuristic to map each test to its focal method.
They search for method names in the test case and map the method mentioned to the test that calls it. They provide the context data as the following: the focal method (FM), focal class name (FC), signatures of the constructor methods of the focal class (C), signatures of other public methods in the focal class (M), and public fields of the focal class (F) -- FM+FC+C+M+F. 
The reason behind these additional contexts is to provide enough information for the model to generate meaningful and compilable tests. 
For example, a test case may need to instantiate an object of the focal class. The model requires the proper context to generate a statement to instantiate them.
Providing the constructor's signature and the focal class's class methods helps the model generate correct instantiating code that does not throw compilation errors. 
Based on their results, this combination of contexts has shown the best results. We have used all the data provided per method to fine-tune our model for test case generation tasks.

\textbf{Defects4j} \cite{just2014defects4j} is a dataset of Java projects consisting of a collection of reproducible bugs and a supporting infrastructure to advance software engineering research. The initial version of \emph{Defects4J} contains 357 real bugs from 5 real-world open-source projects. Each project also has a comprehensive test suite that can expose each bug in each version of the project.
Each version can be accessed using the provided scripts to check out different commits. The updated version of \emph{Defects4j} has 17 projects. In our experiments, we used 5 out of the 17 projects.
Section \ref{sec:evaluation} provides more details on the selection criteria of projects.

\subsection{Pre-Trained Code Model} 
\label{model}
In our study, we have chosen \emph{CodeT5} as our code model for the following reasons: 

Firstly, we did not choose the recent LLMs, e.g., \emph{GPT-4}, \emph{PaLM}, and \emph{LLaMA}, as they are too large to cost-effectively fine-tune downstream tasks and adapt the domain for the project data.
Even though \emph{CodeT5} is much smaller than the recent models, its size gave us a reasonable cost while beating the large LLMs, i.e., \emph{GPT-4}.

Secondly, out of the smaller models, we have chosen \emph{CodeT5}  as it is based on \emph{T5} \cite{raffel2020exploring}. \emph{T5} is a Transformer that uses denoising sequence-to-sequence (Seq2Seq) pre-training and has shown promising results for understanding and generation tasks. \emph{T5}-based models are better for our use case than \emph{BERT} or \emph{GPT}-based models, given that they are not encoder-decoder models. 
Despite their success, \emph{BERT}-based models are encoder-only, and \emph{GPT}-based models are decoder-only architectures.
Encoder models are usually best suited for understanding tasks, and decoder models are suitable for generation tasks. However, the test case generation task requires both understanding and generation skills. 
For instance, prior work that leverages \emph{BERT} for a generation task (code summarization) \cite{feng2020codebert} had to add a separate decoder, which does not benefit from pre-training. 

Lastly, \emph{CodeT5} considers the token type information in the code data. Most other code-aware models use the conventional NLP pre-training techniques. However, code data has rich structural information essential for fully understanding code functionality.

Therefore, given the resources and the application we needed to train our method, we used \emph{CodeT5} as our code model.

\emph{CodeT5} has two main configurations: \emph{CodeT5-small} with 60M parameters and \emph{CodeT5-base} with 220M parameters. 
The maximum source and target sequence lengths are 512 and 256, respectively. 
There are different downstream tasks that \emph{CodeT5} can be fine-tuned on, such as summarize, \emph{CONCODE} (text-to-code generation), translate(Code-to-code translation), code refinement, code defect detection, and code clone detection. 
Two of the tasks mentioned above can be used for test generation, which consists of code and test as input. 
Based on our initial experimental results, the \emph{CONCODE} task generated better test cases than the code-translation task. It seems rational as the code translation model generates code in a different programming language with the same semantics as the original code. However, the input and output do not have the same structure and semantics in test generation. Therefore, in our experiments, we have selected the \emph{CONCODE} sub-task for training the \emph{CodeT5} model with the \emph{CodeT5-base} (the larger) configuration. 
We have used the default hyperparameters for the \emph{CodeT5-base} model. The only change we made in the configuration was the batch size due to the limited GPU memory. We used four as our batch size, the batch data that fit most in our GPU memory. 

\subsection{Baselines}
We chose the following two baselines to compare our results to other methods.
\begin{enumerate}
    \item \textbf{\emph{GPT-4}} is a highly advanced LLM that can mimic human-like speech and reasoning by training on a vast library of existing human communication. It can solve complex problems more accurately, generate creative content, and exhibit human-level performance on various professional and academic benchmarks.
    \item \textbf{\emph{A3Test}} is a test case generation method that exploits domain adaptation at a task level, i.e., test oracle generation, in which they transfer the knowledge learned from oracle generation to a whole unit test case generation.
\end{enumerate}
By adding these two methods as our baseline, we aim to assess how our project-level domain adaptation works compared to the state-of-the-art LLM and a novel test case generation method that exploits a task-level domain adaptation.

\subsection{Evaluation Metrics}
\label{sec:metrics}
In this paper, we use seven different evaluation criteria to evaluate the performance of test generation, i.e., parse rate, execution rate, line-level code coverage, mutation score, adapted mutation score, BLEU score, and CodeBLEU score. 

\subsubsection{Parse and Compile Rate}
{
We report the ratio of parsable and compilable test cases generated by each baseline. We use Tree-sitter \footnote{\url{https://tree-sitter.github.io/tree-sitter/}} parser to evaluate the syntax correctness of the generated test case. Within the parsable test cases, we inject them and compile the project to assess the compilability of the generated test cases.
}

\subsubsection{Line-Coverage and Mutation Score}
\label{Line-Coverage}
We use line coverage and mutation score for the test case adequacy metric.
From a practical point of view, a useful metric for evaluating a test case is a test adequacy metric such as code coverage or mutation score. Other metrics are provided for comparison with baseline literature and their simplicity.  
Although most studies use BLEU score \cite{papineni2002bleu} or CodeBLEU \cite{ren2020codebleu} for evaluating the quality of generated code or test case, {these metrics are sub-optimal for evaluating testing efficacy.} For instance, both BLEU and CodeBLEU are calculating the similarity of the generated output with the ground truth. So, different identifier tokens in the generation will significantly affect its scores while not affecting the functionality of the code or test case.
The model can generate a compilable, meaningful, and effective test case that is not similar to the ground truth. 
Therefore, we picked a simple and basic coverage metric (line coverage) and standard mutation score as our evaluation metrics. In the future, the study can be extended with other adequacy metrics. 

To calculate line coverage, we select only the compilable generated test cases, inject them into the project, and run them.
For RQ1 and RQ2, we calculate the line coverage of the total project, excluding the lines in the test project, i.e., src/main/test.
For RQ3, we calculate the exact code line from the input the generated test case covers.
Clover~\cite{OpenClover} calculates the exact mapping between each test case and its covered lines in the code.
To use Clover, we need to add instrumentation scripts to the build system of each project under study.
Since different projects in \emph{Defects4j} may use other build systems, we included 5 out of 17 projects compatible with Clover.

To emphasize the practical usefulness of our work, we also report mutation score, a fault-based adequacy metric, to demonstrate that our approach can find bugs that the developer-written test, the most related work, and the search-based baseline method can not.
Using their defined mutation operators, we report the standard mutation score (the number of killed mutants divided by the total number of mutants) reported by the Major mutation testing tool. For RQ3, we also report (when applicable) the adapted mutation score, which is the number of killed mutants divided by the covered mutants. 

\subsubsection{BLEU and CodeBLEU} These two metrics calculate the similarity of the generated code compared to the ground truth. Since integrating test adequacy metrics in the training loop is not feasible (it requires execution, which is very costly), we used BLEU to select the best model during the training. We also provide the BLEU and CodeBLEU scores to compare the model before and after domain adaptation.
BLEU calculates the similarity of two texts by calculating their N-Gram co-occurrence.
Since BLEU only evaluates the textual similarity, it is not considered optimal in calculating the similarity of two code snippets.
While BLEU solely calculates the co-occurring n-grams of tokens, CodeBLEU leverages a weighted n-gram to encapsulate different importance of keywords (i.e., $public$, $int$, $return$). It uses syntactic matching via AST and semantic matching via data-flow \cite{guo2021graphcodebert}.

\subsection{Configurations and Environment Setup}
We have used the \emph{CONCODE} downstream task with \emph{CodeT5}-base configuration as our model. We fine-tune the model on the \emph{Methods2Test} dataset and evaluate the results using \emph{Defects4j} projects and \emph{EvoSuite}-generated tests.
All the model hyperparameters are set as the default for the \emph{CONCODE} configuration of \emph{CodeT5}. The batch size is set to 4, the maximum that can fit our setup's GPU memory. We fine-tune all the model layers for 20 epochs for the domain adaptation step. 

We have used a single ComputeCanada (Beluga) node for all experiments, with 4 32GB V100 GPUs, 10 CPU cores, and 80GB RAM. However, with minor changes in batch size and training time, all experiments can be executed with 16GB GPU, 1 CPU core, and 10 GB RAM. 



\begin{table*}[ht!]
\caption{Evaluation metrics scores for \emph{CodeT5} without Domain Adaptation (DA) (RQ1). Comparison of \emph{Code-T5} with DA) versus \emph{GPT-4} and \emph{A3Test} is also shown for RQ2. Bold values denote the best metric score for each project compared to the baselines.}
\label{tab:rq1n2}
\resizebox{0.8\textwidth}{!}{
\begin{tabular}{|c|c|c|c|c|c|c||c|}
\hline
\textbf{Baselines} & \textbf{Metrics} & \textbf{compress} & \textbf{gson} & \textbf{jksnCore} & \textbf{jksnDB} & \textbf{jsoup} & \textbf{AVG} \\ \hline
 & { Parse Rate} & { 20.75} & { 24.01} & { 14.21} & { 18.26} & { 39.27} & { 23.30} \\ \cline{2-8} 
 & { Compile Rate} & { 1.66} & { 3.67} & { 0.70} & { 0.92} & { 22.51} & { 5.89} \\ \cline{2-8} 
 & BLEU & 11.59 & 18.64 & 16.39 & 18.34 & 25.56 & 18.10 \\ \cline{2-8} 
 & CodeBLEU & 9.15 & 16.64 & 16.98 & 16.78 & 22.10 & 16.33 \\ \cline{2-8} 
 & Line Coverage & 2.00 & 25.60 & 2.10 & 31.40 & 63.10 & 24.84 \\ \cline{2-8} 
\multirow{-6}{*}{\textbf{\begin{tabular}[c]{@{}c@{}}CodeT5\\ without\\ DA\end{tabular}}} & { Mutation Score} & { 0.55} & { 12.26} & { 0.07} & { 11.95} & { 32.78} & { 11.52} \\ \hline \hline
 & { Parse Rate} & { 89.29} & { \textbf{100.00}} & { 93.33} & { 94.46} & { \textbf{100.00}} & { 95.42} \\ \cline{2-8} 
 & { Compile Rate} & { \textbf{39.29}} & { \textbf{47.67}} & { \textbf{38.33}} & { \textbf{28.37}} & { \textbf{62.50}} & { \textbf{43.23}} \\ \cline{2-8} 
 & BLEU & \textbf{40.84} & \textbf{42.06} & \textbf{28.41} & \textbf{36.74} & \textbf{44.36} & \textbf{38.48} \\ \cline{2-8} 
 & CodeBLEU & \textbf{22.37} & \textbf{35.12} & \textbf{30.06} & \textbf{31.70} & \textbf{44.10} & \textbf{32.67} \\ \cline{2-8} 
 & Line Coverage & \textbf{32.80} & \textbf{52.20} & \textbf{21.20} & \textbf{43.10} & \textbf{68.00} & \textbf{43.46} \\ \cline{2-8} 
\multirow{-6}{*}{\textbf{\begin{tabular}[c]{@{}c@{}}CodeT5\\ with\\ DA\end{tabular}}} & { Mutation Score} & { \textbf{20.53}} & { \textbf{35.61}} & { \textbf{8.70}} & { \textbf{28.60}} & { \textbf{46.42}} & { \textbf{27.97}} \\ \hline
 & { Parse Rate} & { \textbf{99.28}} & { 98.55} & { \textbf{98.37}} & { \textbf{98.08}} & { 99.40} & { \textbf{98.74}} \\ \cline{2-8} 
 & { Compile Rate} & { 2.90} & { 17.15} & { 4.20} & { 7.52} & { 22.75} & { 10.90} \\ \cline{2-8} 
 & BLEU & 18.53 & 26.39 & 18.29 & 22.43 & 27.11 & 22.55 \\ \cline{2-8} 
 & CodeBLEU & 18.73 & 28.19 & 23.32 & 23.87 & 25.65 & 23.95 \\ \cline{2-8} 
 & Line Coverage & 0.70 & 32.40 & 4.10 & 33.20 & 56.80 & 25.44 \\ \cline{2-8} 
\multirow{-6}{*}{\textbf{GPT-4}} & { Mutation Score} & { 0.10} & { 15.93} & { 0.98} & { 14.44} & { 43.45} & { 14.98} \\ \hline
 & { Parse Rate} & { 53.70} & { 64.61} & { 44.98} & { 68.16} & { 56.25} & { 57.54} \\ \cline{2-8} 
 & { Compile Rate} & { 1.93} & { 6.85} & { 1.27} & { 1.07} & { 17.14} & { 5.65} \\ \cline{2-8} 
 & BLEU & 11.33 & 16.44 & 13.08 & 15.75 & 18.79 & 15.08 \\ \cline{2-8} 
 & CodeBLEU & 7.42 & 15.61 & 13.32 & 15.58 & 18.11 & 14.01 \\ \cline{2-8} 
 & Line Coverage & 2.00 & 29.50 & 2.00 & 31.60 & 52.80 & 23.58 \\ \cline{2-8} 
\multirow{-6}{*}{\textbf{A3Test}} & { Mutation Score} & { 0.00} & { 12.85} & { 0.01} & { 11.95} & { 34.98} & { 11.96} \\ \hline
\end{tabular}}
\end{table*}

\section{Results and Analysis}
\label{sec:evaluation}
\subsection{RQ1: Effectiveness of \emph{CodeT5} without DA}
\label{sec:rq1_result}

\noindent \textbf{Experiment Design.} 
First, we explain how we split \emph{Defects4j} data into training and evaluation sets per project.
In RQ1, we only use the evaluation set to assess the base model \emph{CodeT5} without Domain Adaptation (DA). 

There are two ways to split the train and evaluation set on \emph{Defects4j}.
We can randomly select 20\% of the lines and move the line-test tuples to the evaluation set.
The problem with this approach is that we may have a data leak between the train and the evaluation set.
For example, in a method with five lines, each line is mapped to a test.
In some cases, all five lines are mapped to the same test case. If we randomly pick 2 of 5 lines for the evaluation set, the model will have access to the other three lines, which consist of the same output test that we expect the model to generate in the evaluation.
Since this constitutes a data leak, we divided the data at the test case level.

In this approach, a leave-one-out evaluation \cite{kocaguneli2013software}, we first make a set of all unique test cases in the dataset per project. We randomly select 20\% of them for the evaluation set per project.
Finally, the evaluation set is created using the line-test tuples of those 20\% test cases. 
This way, the dataset will not have identical test cases between the evaluation and training sets, i.e., no data leaks.
Note that some test cases might still cover some lines in the evaluation set in the training test.
However, those test cases in the training set are not the same as the main test cases in the test set, which was selected in the 1-to-1 mapping procedure for that given line, which the model tries to generate.
We do not consider the following a data leak since the target test cases generated will no longer be the same as those seen in the training. 

For RQ1, we investigate the ability of \emph{Code-T5} to generate test cases by only applying fine-tuning on downstream tasks.
We use the \emph{Methods2test} dataset to fine-tune the test generation task.
After training the model on the \emph{Methods2test} dataset, we directly evaluate the model on the evaluation set split, as mentioned above for splitting \emph{Defects4j}.
To generate test cases for \emph{Defects4j} projects, we need to extract a context similar to the structure of the \emph{Methods2test} dataset.
After generating the tests, we calculate the line coverage on each project. 
The line coverage is the number of covered lines divided by the total number of lines in the src/main/java folder.
The line coverage on the \emph{CodeT5} without DA baselines shows the line coverage by the tests generated by the model without domain adaptation on the evaluation project.
{We only use test cases that pass for calculating the mutation scores, as we need a green test suite to set up the mutation testing process.}

\noindent \textbf{Results.}
As demonstrated in Table \ref{tab:rq1n2}, we calculate the evaluation metrics of model-generated tests on five Defects4j projects using \emph{CodeT5} without DA.

We have two findings in this RQ: 
{(a) In most cases, the model-generated test cases were not compilable due to the inability to infer the correct dependencies, which led to the generation of undefined objects.
Also, the model generated truncated test cases to the limited output length per sample in \emph{CodeT5} (512 tokens).
(b) Existing studies such as \cite{tufano2020unit} only reported the coverage (a high coverage in this case for a small (18) set of sample methods), which is not representative of the actual quality of the model.
Most other studies only report generic static metrics, such as the BLEU score, which fails to capture the test adequacy. 
However, our results revealed that the generated test cases' test adequacy metrics (line coverage and mutation scores) are not as promising as the BLEU or CodeBLEU.
They are also much lower than the reported coverage in the original \emph{AthenaTest} paper for the 18 small sample codes they have assessed.}

\begin{tcolorbox}[
    width=8.5cm,
    boxsep=0pt,
    left=3pt,
    right=3pt,
    top=3pt,
]
\textbf{Answer to RQ1:}
{The results of \emph{CodeT5} without DA show that fine-tuning with only task-specific data is insufficient to generate test cases that are compilable or test-adequate.}
\end{tcolorbox}

\subsection{RQ2: Effectiveness of CodeT5 with DA}
\label{sec:rq2_result}

\noindent \textbf{Experiment Design.} 
In this subsection, we explain the details of the experiment procedure used in RQ2.

We use the same splits of Defects4j as mentioned in the previous RQ.
To apply domain adaptation, we use the training set of Defects4j to train the fine-tuned model, i.e., \emph{CodeT5} with DA.
Then, we generate test cases using the same evaluation set to calculate the metric scores.

We have two state-of-the-art baselines to compare our approach: \emph{GPT-4} and \emph{A3Test}. 
To generate test cases by \emph{GPT-4}, we had to develop a new style of feeding the input, as our dataset has a lot of redundancy due to its granularity being at the line level.
Asking the model to create tests for each line individually wastes resources.
To mitigate this issue and optimize prompting cost (to make this solution more practical), we ask \emph{GPT-4} to create as many tests as it needs to maximize line coverage for a given method.

The prompt template is shown in Listing \ref{lst:prompt}.
First, we query the model with a system prompt that defines the model's role, a unit test case generator with meaningful assertions (a nontrivial requirement for generating unit test cases \cite{watson2020learning}).
For the actual task, we provide the focal method and its context and ask the model to generate a unit test case that covers the maximum line coverage for the focal method.
We let the model generate as many tests as it needs (since other baselines of comparisons create multiple tests as well), but to avoid redundant tests, we ask to generate new tests only if they cover new lines of code.
We also added minor instructions to the prompt to make our post-processing easier, i.e., only generating Java code, using [TCS] tokens to separate multiple test cases, removing natural language comments and @Test annotations, and substituting new lines with [EOL] tokens.

For \emph{A3Test}, we use their already fine-tuned model and their testing script provided in their replication package.
Since \emph{A3Test} is also a token-to-token generation model that receives a structure similar to ours in the input, i.e., focal method + focal context.
We use the same evaluation set splits from our Defects4j dataset (from RQ1).
All the hyper-parameter settings were used according to what was reported in their paper or the default values suggested in their replication package.


\lstset{basicstyle=\ttfamily\footnotesize,breaklines=true}
\begin{figure}[t!]
\centering
\begin{lstlisting} 
prompt = [
    {"role": "system",
     "content": f"You are a unit test case generator
      with meaningful assertions for Java project: {prj}."},
    {"role": "user", "content": f"""Given a focal method
      surrounded by ???, generate unit test case methods
      that cover maximum line coverage. Only create new
      tests if they cover new lines of code. Only generate
      the Java code part of test methods. Use [TCS] to
      separate the multiple test cases. Input text:
      ???{method}???"""},
    {"role": "user", "content": """Remove all comments
      (e.g. line starts with // and surrounded by /* and */), 
      NL description and @Test annotations. New lines
      should be substituted with [EOL]."""}
]
\end{lstlisting}
\vspace{-0.1in}
\caption{Prompt used for GPT-4}
\label{lst:prompt}
\end{figure}


\noindent \textbf{Results.} 
Table \ref{tab:rq1n2} reports the evaluation metrics to compare \emph{CodeT5} with DA and the studied baselines.
\emph{CodeT5} without DA refers to the \emph{CodeT5} model fine-tuned on test generation downstream task using the \emph{Methods2test} dataset, without any domain adaptation. 
\emph{CodeT5} with DA refers to the model after applying project-level domain adaptation.
The results show that using project-specific data for domain adaptation significantly increases the model's performance.
The average improvement of percentage points over all projects is {72.12\% for parse rate, 37.34\% for compile rate}, 20.38\% for BLEU, and 16.34\% for CodeBLEU, 18.62\% for line coverage, {and 16.45\% for mutation score}.

The results of test adequacy metrics are new and promising.
As discussed, most related work does not report these metrics due to the effort required to make all test cases executable.
Our results reveal that without project-specific domain adaptation, the metrics are low, with line coverage between 2\% to 63\%, with a median of 25.6\% and a mean of 24.84\%. {For mutation score, it ranged between 0.07\% to 32.78\%, with a median of 11.95\% and a mean of 11.52\%}.
However, the metrics improve significantly after applying the domain adaptation, with line coverage to a range between 21.20\% and 68\%, a median of 43.10\%, and a mean of 43.46\%. {For the improved mutation score, it ranged between 8.70\% and 46.42\%, with a median of 28.60\% and a mean of 27.97\%}. 
In other words, there was a 17.5\% improvement over the median and an 18.62\% improvement over the mean for line coverage; 16.65\% improvement over the median, and 16.45\% over the mean for mutation score in percentage points. From the result, we can observe that applying domain adaptation to transfer project-specific knowledge has a substantial improvement in unit test generation, both in static textual similarity and test adequacy metrics. The reason could be that unit test generation heavily relies on the internal knowledge of the software under test. Just tuning the models at a task level without enough knowledge of the software will have a marginal effect on the testability of the generated unit tests.


We also compare our approach with two state-of-the-art baselines, i.e., \emph{GPT-4} and \emph{A3Test}.
As shown in Table \ref{tab:rq1n2}, none of the baselines could outperform \emph{CodeT5} with DA in all metrics {except for the parse rate of \emph{GPT-4}}.
\emph{GPT-4} showed better overall performance than \emph{A3Test} in all metrics.
{\emph{A3Test} and \emph{CodeT5} without DA showed the least performance.
\emph{A3Test} had the lowest performance in compile rate, line coverage, BLEU, and CodeBLEU. \emph{CodeT5} without DA showed the least parse rate and mutation score performance.
The performance difference between the two least-performing baselines was not very big.
However, the difference between their parse rate scores was significant, with +34.24\% points for \emph{A3Test}.
The results suggest that transferring Oracle generation knowledge has a positive impact in generating syntactically correct test cases.}


One interesting observation is that CodeBLEU scores are relatively higher than BLEU for \emph{GPT-4}.
Even though it generates different n-gram tokens than the ground truth, the AST-matching and dataflow matching scores are relatively higher.
Even though \emph{GPT-4} generates different tokens, e.g., different identifier names, it generates similar code in terms of syntax (AST) and semantics (dataflow).
Generating different identifier tokens can be an inherent trait of \emph{GPT-4} as it is trained on a much more diverse dataset with much larger parameters.
{However, there is more than one way of naming an identifier in source code.}
What determines the function of a code is its syntax and semantics.
{Also, \emph{GPT-4} shows the best performance in parse rate, meaning that it generates the most syntactically correct test cases. 
However, they could not beat \emph{CodeT5} with DA in other metrics as they lack the domain-specific knowledge to make the code locally correct for compilation and effective test adequacy.}
From this observation, we foresee good potential on \emph{GPT-4} if paired with the proper tuning strategy, e.g., prompt-tuning, fine-tuning, domain adaptation, etc.


\begin{tcolorbox}[
    width=8.5cm,
    boxsep=0pt,
    left=3pt,
    right=3pt,
    top=3pt,
]
\textbf{Answer to RQ2:}
{Overall, the results suggest that applying project-specific domain adaptation improves \emph{CodeT5} by \textbf{72.12\%} in parse rate, \textbf{37.34\%} in compile rate, \textbf{20.38\%} in BLEU, \textbf{16.34\%} in CodeBLEU, \textbf{18.62} in line coverage, and \textbf{16.45\%} in mutation score over the one without DA. 
It also significantly outperforms all the other baselines, except for the parse rate of \emph{GPT-4}.}
\end{tcolorbox}

\subsection{{RQ3: Augmentation with SBST}}
\label{eval:rq3}

\noindent \textbf{Experiment Design.}
In RQ3, we compare our approach with \emph{EvoSuite}, a well-known search-based approach for test case generation.
First, we run \emph{EvoSuite} with the default settings (10 minutes per class) to generate tests for all classes in the project.
We use the same train-test split as RQ1 and RQ2.
We calculate the line coverage and the mutation scores using the \emph{EvoSuite} framework.
The purpose of our proposed framework is not to compete with or replace \emph{EvoSuite} but to complement or augment such existing approaches.
Therefore, we also report the number of ``new lines'' our test cases can cover compared to what was covered already by the \emph{EvoSuite}-generated test suite.
Also, note that these ``new lines'' are not covered by the developer-written test cases of the training sets either. Thus, the study emphasizes the tool's impact by comparing existing automated testing tools and manual test generation practices. 

\begin{table}[t!]
\caption{Line Coverage comparison between \emph{EvoSuite} and model generated tests. The NewCL has covered lines that neither \emph{EvoSuite} nor the developer-written tests from the training set have covered.}
\label{tab:evo_lc}
\centering
\vspace{-0.1in}
\resizebox{1\linewidth}{!}{
\begin{tabular}{|c|cc|cc|cc|c|}
\hline
Project  & \multicolumn{2}{c|}{Model CL}     & \multicolumn{2}{c|}{EvoSuite CL} & \multicolumn{2}{c|}{New CL}         & Total Lines \\ \hline
compress & \multicolumn{1}{c|}{216}   & 58\% & \multicolumn{1}{c|}{87}   & 23\% & \multicolumn{1}{c|}{174}  & 46.70\% & 372         \\ \hline
gson     & \multicolumn{1}{c|}{458}   & 69\% & \multicolumn{1}{c|}{539}  & 82\% & \multicolumn{1}{c|}{31}   & 4.70\%  & 657         \\ \hline
jksnCore & \multicolumn{1}{c|}{399}   & 30\% & \multicolumn{1}{c|}{674}  & 51\% & \multicolumn{1}{c|}{82}   & 6.20\%  & 1307        \\ \hline
jksnDB   & \multicolumn{1}{c|}{1357}  & 50\% & \multicolumn{1}{c|}{136}  & 5\%  & \multicolumn{1}{c|}{1246} & 48\%    & 2595        \\ \hline
jsoup    & \multicolumn{1}{c|}{192}   & 82\% & \multicolumn{1}{c|}{39}   & 16\% & \multicolumn{1}{c|}{157}  & 66.50\% & 519         \\ \hline \hline
AVG      & \multicolumn{1}{c|}{524.4} & 58\% & \multicolumn{1}{c|}{295}  & 35\% & \multicolumn{1}{c|}{338}  & 34.42\% & 1090        \\ \hline
\end{tabular}}
\end{table}

\noindent \textbf{Results}. 
Table \ref{tab:evo_lc} reports the line coverage of the test cases generated by \emph{EvoSuite} and our framework.
The model-covered lines (Model CL) column shows the number and percentages of lines of code covered by model-generated tests.
\emph{EvoSuite}-covered lines (\emph{EvoSuite} CL) show the number and percentages of lines covered by \emph{EvoSuite}.
Finally, the new covered lines (New CL) column shows the extra lines covered by model-generated tests that \emph{EvoSuite} can not cover.
Note that these lines are not covered by the developer-written test cases of the training sets either.

The results indicate that \emph{EvoSuite} line coverage is higher than our framework in 2 projects, and ours is higher in 3 projects.
Overall, \emph{EvoSuite}'s median and mean coverage are 23\% and 35.4\% vs. ours, which are 58\% and 57.8\%.
However, the motivation of our work is to augment existing test generation systems and not to replace them.
If our model generates tests that can cover new uncovered lines compared to \emph{EvoSuite}, we say our model augments \emph{EvoSuite}; the total coverage will be more than both individually. 
Looking at the New CL column, we see that in 5 out of 5 projects, we can augment \emph{EvoSuite} by covering extra lines.  

\begin{table}[t!]
\caption{Mutation and adapted mutation scores for model generated (Model MS and Model AMS) and \emph{EvoSuite} (Evo MS and Evo AMS) tests. The New MK column shows the number of mutants not killed by \emph{EvoSuite} but by model-generated tests. }
\label{tab:evo_ms}
\centering
\vspace{-0.1in}
\resizebox{1\linewidth}{!}{
\begin{tabular}{|c|c|c|c|c|cc|}
\hline
Project  & Model MS & Evo MS  & Model AMS & Evo AMS  & \multicolumn{2}{c|}{New MK}        \\ \hline
compress & 0.00\%   & 55.90\% & 0.00\%    & 69.50\%  & \multicolumn{1}{c|}{0}   & 0\%     \\ \hline
gson     & 13.50\%  & 64.90\% & 50.00\%   & 100.00\% & \multicolumn{1}{c|}{0}   & 0\%     \\ \hline
jksnCore & 14.80\%  & 87.20\% & 50.70\%   & 100.00\% & \multicolumn{1}{c|}{0}   & 0\%     \\ \hline
jksnDB     & 22.40\%  & 0.00\%  & 54.20\%   & 0.00\%   & \multicolumn{1}{c|}{26}  & 22.40\% \\ \hline
jsoup    & 32.00\%  & 0.00\%  & 47.10\%   & 0.00\%   & \multicolumn{1}{c|}{8}   & 32\%    \\ \hline \hline
AVG      & 16.54\%  & 41.60\% & 40.40\%   & 53.90\%  & \multicolumn{1}{c|}{6.8} & 11\%    \\ \hline
\end{tabular}}
\end{table}

Table \ref{tab:evo_ms} reports the mutation score of model-generated unit tests compared to the \emph{EvoSuite} tests.
We used defects4j to calculate the mutation score.
The low mutant coverage of our approach is because we are using only 20 percent of each project as our test set, but mutants are everywhere. Since the data is divided in a line-level manner and \emph{EvoSuite} needs the whole class for test generation, we could not use \emph{EvoSuite} to only generate tests for the test set portion of the dataset (the portion that was given to the trained model for test generation).
So, a direct comparison is not straightforward.

The same problem also exists in the coverage calculation since \emph{EvoSuite} generates a test suite for the whole project.
However, we could select only the lines in our test set for coverage comparison using Clover coverage reports, which we cannot do with the mutation tool.
To better reflect the mutation-killing power of our approach, we calculated the Adapted Mutation Score, which compares the model's ability to kill the covered mutants.

Comparing the two techniques, the mutation score of model-generated tests is higher than \emph{EvoSuite} for 2 (Jacksondatabind, Jsoup) out of 5 projects.
For example, in jksnDB, we kill 26 new mutants.

As mentioned, all the above results for \emph{EvoSuite} are collected with the default \emph{EvoSuite} setup, which is 10 minutes timeout per class.
One can argue that \emph{EvoSuite} might generate more new lines if we set a higher time.
Although this is true in theory, first, the default values are chosen based on hyper-parameter tuning, which means that, on average, one won't get much more coverage by simply giving more time for test generation per class. Second, we also noticed that most projects would converge even before 10 minutes.

As explained before, we recommend using our approach in addition to a tool like \emph{EvoSuite}.
Our suggested use case in practice is to start with \emph{EvoSuite} (with a default budget).
Then, identify lines not covered by \emph{EvoSuite} and pass them to our framework to generate test cases instantly. 
Note that one test case generation in our framework takes around 2 seconds (including all post-processing steps) compared to 153 seconds per test case on average for \emph{EvoSuite} on these projects.
Our approach provides a fast add-on to \emph{EvoSuite}, especially for new commits, since otherwise, one would need to rerun \emph{EvoSuite} for the whole class with every minor change.  

Finally, note that in addition to improving performance and being faster than the alternative search-based approach, the other attraction of our work is to focus on generating readable and more maintainable test cases, given that they are derived based on developer-written test cases rather than predefined templates of search-based approaches.
Although we did not study this aspect in detail in this paper and only showed an example \ref{fig:test_sample}, our baseline paper \cite{tufano2020unit} provides some evidence based on their user study.

\begin{tcolorbox}[
    width=8.5cm,
    boxsep=0pt,
    left=3pt,
    right=3pt,
    top=3pt,
]
\textbf{Answer to RQ3:}
In general, our approach can increase the coverage and mutation score of the existing state-of-the-art test generation techniques such as \emph{EvoSuite} and thus is recommended to be used together with such tools.
\end{tcolorbox}

\begin{figure}[t!]
\centering
\begin{lstlisting} 
public void testHashCode1609() {
    ArcDialFrame f1 = new ArcDialFrame();
    ArcDialFrame f2 = new ArcDialFrame();
    assertTrue(f1.equals(f2));
    int h1 = f1.hashCode();
    int h2 = f2.hashCode();
    assertEquals(h1, h2);
}
\end{lstlisting} 
\vspace{-0.1in}
\caption{An example of model-generated tests.}
\label{fig:test_sample}
\end{figure}
\section{Limitations and threats to validity}
\label{sec:threats}
One of the limitations of our approach is that it might depend on developer-written test cases.
Although training our model on automatically generated tests is possible, it could hinder the benefits we were targeting, such as better fault detection and readability.
Therefore, our approach's use case is to extend existing tests so that new lines are covered, and new faults are detected.
However, in practice, this is not a considerable hindrance since, except for newly created projects, most reasonable projects come with some tests already in their regression test suite. Thus, our approach can start with those test suites and improve and augment them. Alternatively, suppose the project does not have any test cases. In that case, the next best option is using automatically generated test cases that are generated by an LLM such as \emph{GPT-4} so that the initial tests are still readable and have a relatively high quality, to begin with.

Regarding construct validity threats and the effectiveness of the metrics, we made sure we went beyond code coverage and looked at the mutation score. New mutants killed by our approach mean potentially new faults can be detected by the model-generated tool compared to what the developers have detected. 
However, we did not provide a systematic study for the readability of test cases and only showed an example. We neglected this part since the baseline paper \cite{tufano2020unit} already has done a user study and reported the readability as a benefit of model-generated tests. 

{
Regarding the threat to external validity, one closely related study we failed to compare with was \emph{ChatTester}. We couldn't properly run their tool on the five Defects4j projects used in this study, mainly due to their data-pair collection component. The component collected 20 pairs of data instances from one project and no pairs for the other four projects, which was insufficient for comparison.
To mitigate this, we compared our work with \emph{GPT-4} to investigate the test case generation performance of state-of-the-art LLM. Since in our comparison, we did not employ advanced prompt engineering strategies such as those used on \emph{ChatTester}, future studies are needed to compare our work with advanced prompt-engineered LLMs for test generation.
}

{
Also, we agree that the selected projects can threaten this study's external validity. However, Defects4j is a well-known and widely used dataset with quality unit test cases. Using a limited number of projects is mainly due to the considerable resource cost of calculating the test adequacy metrics. Due to the same reason, some of the previous studies that evaluate test adequacy metrics (i.e., line coverage or mutation score) on test or test oracle generation also experimented with a small number of projects like us, e.g., TOGA \cite{dinella2022toga} and A3Test \cite{alagarsamy2023a3test}. Also, it is worth noting that running GPT-4 per each extra project is very costly, hindering the experiments' size.
}
\section{Conclusion}
\label{sec:con}
This study showed that code models can be fine-tuned on test generation downstream tasks. However, their performance is ineffective on a new project compared to search-based approaches due to domain shift. To mitigate the problem, we proposed a domain adaptation framework that leverages existing developer-written tests. We showed that applying project-level domain adaptation improves the quality of the generated test cases w.r.t. compilability, similar to human-written and test-adequate. Our approach outperforms the largest state-of-the-art LLM, \emph{GPT-4} on all metrics except the parse rate and all metrics for \emph{A3Test}, the deep test case generation method that exploits task-level domain adaption.
Finally, we compared our proposed framework with state-of-the-art search-based approaches and showed that our approach could complement and increase line coverage and mutation score. 
In the future, we will explore other code models and expand the experiment on new datasets.
We will also run our user study to evaluate better the generated tests' readability.

\begin{acks}
This work was partially supported by the NSERC Discovery Grant (RGPIN/04552-2020), and the NSERC and Alberta Innovates Alliance Grant (ALLRP/568643-2021). 
\end{acks}

\balance
\bibliographystyle{ACM-Reference-Format}
\bibliography{main}


\end{document}